\documentclass[12pt,preprint]{aastex}

\slugcomment{}

\shorttitle{Radio Emission from LP 349-25}
\shortauthors{Osten et al.} 

\begin{document}
\title{Steady and Transient Radio Emission from Ultracool Dwarfs } 
\shorttitle{Steady and Transient Radio Emission from Ultracool Dwarfs } 
 
\author{Rachel A. Osten\altaffilmark{1,2}}
\affil{Astronomy Department, University of Maryland, College Park MD 20742 }
\email{osten@stsci.edu}
\author{N. Phan-Bao}
\affil{Institute of Astronomy and Astrophysics, 
Academica Sinica, P. O. Box 23-141, Taipei 10617, Taiwan R.O.C.}
\email{pbngoc@asiaa.sinica.edu.tw}
\author{Suzanne L. Hawley}
\affil{Astronomy Department,  Box 351580, University of Washington, Seattle, WA 98195}
\email{slh@astro.washington.edu}
\author{I. Neill Reid}
\affil{Space Telescope Science Institute, 3700 San Martin Drive, Baltimore, MD 21218}
\email{inr@stsci.edu}
\author{Roopesh Ojha}
\affil{NVI/United States Naval Observatory, Washington, DC 20392-54200}
\email{rojha@usno.navy.mil}
\altaffiltext{1}{Hubble Fellow}
\altaffiltext{2}{Currently at Space Telescope Science Institute, 3700 San Martin Drive, Baltimore, MD 21218}

\begin{abstract}
We present the results of multi-frequency radio observing campaigns designed to elucidate the nature of radio
emission from very low mass stars.
We detect radio emission in an additional two epochs of the ultracool dwarf binary LP 349-25, finding 
that the observed emission is broadband and steady on timescales between 10s and 10.7 hours, as well as on timescales of 0.6 and 1.6 years.
%The lack of variability coupled with 
%undetectable amounts of circular polarization do not support a coherent emission mechanism.
This system is unusual for ultracool dwarfs with detectable radio emission, in exhibiting a lack
of any large scale variability, particularly the bursting (periodic or aperiodic) behavior exhibited
by the other objects with detectable levels of radio emission.
We explore the constraints that the lack of variability on long and short timescales, and flat spectral
index, imply about the radio-emitting structures and mechanism.  
The temporal constraints argue for a high latitude emitting region with a large inclination so that it
is always in view, and survives for at least 0.6 year.
Temporal constraints also limit the plasma conditions, implying that the electron density
be $n_{e} <$4$\times$10$^{5}$ cm$^{-3}$ and $B<$130 G in order not to see
time variations due to collisional or radiative losses from high energy particles.
The observations and
constraints provided by them are most compatible with a nonthermal radio emission mechanism,
likely gyrosynchrotron emission from a spatially homogeneous or inhomogeneous source.
This indicates that, 
similar to behaviors noted for chromospheric, transition region, and coronal
plasmas in ultracool dwarfs,
the magnetic activity patterns observed in active higher mass stars can survive to the 
substellar boundary.
%The characteristics of the radio emission (flux density, spectral index, polarization) from LP 349-25
%fit with the quiescent emission derived for active stars, and suggest that gyrosynchrotron emission
%is the dominant radio-emitting mechanism.  
%VLA observations at higher spatial resolution do not 
%provide constraints on which star is the source of the radio emission, although future 
%observations with the High Sensitivity Array will.
%This steady, broad-band behavior appears to indicate a new type of phenomenon in ultracool dwarf radio
%emission.
We also present new epochs of multi-frequency radio observations for the the ultracool dwarfs 2MASS 05233822-140322 and
2MASS14563831-2809473(=LHS~3003); each has been detected in at least one previous epoch but are not detected in
the epochs reported here.  
%LHS~3003 is the second ultracool dwarf to display sporadic radio emission.
The results here suggest that magnetic configurations in ultracool dwarfs can be long-lasting,
and support the need for further radio monitoring using a simultaneous, multi-frequency observing
approach.
%The results here support the need for simultaneous multi-frequency recordings of radio emission from these enigmatic
%objects.
\end{abstract}


\section{Introduction}
It was initially expected that radio emission would not be detected in ultracool dwarf stars, given the
evidence that the number of objects exhibiting magnetic activity past a spectral type of M7V
was declining \citep{hawley1996,west2004} despite increasingly rapid rotation; this seemed to match expectations that the cool, dense, and
globally neutral atmospheres \citep[fractional ionizations $\le$ 10$^{-5}$;][]{mohanty2002}
of ultracool dwarfs would be unable to sustain the kinds
of magnetic stresses thought to operate in the atmospheres of solar-like dwarf stars 
\citep[spectral types F, G, K, and early M;][]{mohanty2002}.
\citet{reinersbasri2007}
 showed
that large-scale fields can be maintained by detecting their effect on transitions in molecular absorption bands,
but the fields' connection to the non-radiative heating
is still questionable.

Since the discovery of radio emission from
one of these objects \citep{berger2001}, there have been two key findings: the overall number
of radio-emitting ultracool dwarfs is relatively low, $\sim$10\% \citep{berger2006} with a slight
preponderance for objects of spectral type M9 to more commonly display detectable levels of
radio emission \citep{phanbao2007}; and there is a wide variety of behavior exhibited by
those objects which have shown detectable levels of radio emission. Several objects have
displayed pulsar-like behavior \citep{berger2005,hallinan2006} with periods consistent
with those inferred from rotational periods using radius and $v\sin i$ measurements;
interpulse emission is observed but seems to be less circularly polarized.
A few others show large increases in flux density, accompanied by large amounts of circular
polarization \citep{burgasserputman2005} while still others have been detected in one epoch
and undetected in others \citep{antonova2007}. 
Variability is at the core of these behaviors; even the same object observed multiple times
at radio wavelengths shows disparate behaviors \citep{berger2002,hallinan2006,hallinan2007,berger2008}.

As with higher mass stellar radio emission \citep{osten2007} it is possible that more than one type of radio emission
mechanisms could be operating in the coronae of very low mass dwarfs: one, associated with 
gyrosynchrotron emission, is incoherent and consists of steady and variable wide-band
radio emission, with moderate circular polarizations and roughly flat spectra; the other type being a coherent mechanism,
possibly cyclotron maser emission, which can be extremely variable, usually highly circularly polarized,
and can have steep spectra.  
\citet{ostenetal2006} attributed the steady radio emission
from one ultracool dwarf to gyrosynchrotron emission based on its spectral index and circular polarization
behaviors.
The dramatic radio variability which has been observed on a few very low mass objects generally includes
highly circularly polarized emission and points to a different radio
emission mechanism than usually invoked for higher mass stellar radio emission, probably coherent.
Current speculations are that it may be cyclotron maser emission \citep{hallinan2007}. 
Still unknown is how the source population of accelerated particles which produces this emission
originates.
%while \citet{hallinan2006,hallinan2007} have interpreted the periodic variations seen in a few
%ultracool dwarfs to cyclotron maser emission, in analogy with solar system giant planet radio emission.

Although transition region and X-ray observations detect persistent
fluxes from very low mass stars at levels compatible with scaling laws (based on H$\alpha$) extrapolated from earlier
type M dwarfs, very low mass dwarfs appear to violate another tight relationship
between magnetic activity indicators, namely the X-ray luminosity -- radio luminosity
correlation pointed out by \citet{gudelbenz1993} and \citet{benzgudel1994}.  
The detection of both persistent and flaring radio emission from a handful
of late M and L objects, reported in \citet{berger2001} and \citet{berger2006}, suggests
that nonthermal radio emission is much stronger than predicted in some objects at late spectral types.  
The interpretation of the radio emission, however, is complicated by the fact that
more than one potential mechanism may be at work: on the Sun and on other active
stars, both incoherent and coherent nonthermal radio emissions
produce detectable levels of radio emission \citep[see, e.g.][]{dulk1985,gudel2002}.
Detailed observations of solar radio emission show that a variety of emissions can be produced in the
centimeter wavelength range depending on the type and location of emitting structures and their variation with time.
Indeed, this provides a simple
resolution to the apparent ``violation'' of the L$_{X}$--L$_{R}$ relation in very low
mass dwarfs noted by, e.g. \citet{berger2005}.

Studying stellar radio emission
is important because it is the only wavelength region in which direct evidence of the role of accelerated electrons in stellar atmospheric structures and dynamics can
routinely and easily be seen.  In order to make advances in knowledge about atmospheric structure and dynamics, temporal and
spectral information is needed.  Detection epochs typically concentrate on building up exposure time at a single
frequency with on-source times sufficient only to see very short time-scale variability such as bursting
behavior.  Longer on-source times can probe rotationaly-induced variability as well as increase the
likelihood of observing flares.
%; current flare duty cycle estimates are 5--7\% (ref.)
Simultaneous multi-frequency radio observations are better at pinning down the emission mechanism
and allowing for spectral-temporal modelling to better understand the emitting structures. To this end, 
we selected several objects which had been previously detected at a single radio frequency for follow-up with
multi-frequency observations.  The properties of the program sources are summarized in Table~\ref{tbl:sources}.
The paper is organized as follows: \S 2 describes the properties of the program sources, \S 3 describes the
observations, \S 4 discusses the data analysis and results, \S 5 provides a general discussion on how these
observations contribute to our understanding of the nature of radio emission from very low mass stars,
and \S 6 concludes what these data reveal about radio emission from ultracool dwarfs.

\section{Program Sources}
\subsection{LP349-25}
\citet{gizis2000}
recognized LP349-25 as a nearby star, deriving a spectral type of M8V with an estimated photometric distance of 8.3 pc. 
\citet{close2002} observed the system with
adaptive optics in an earlier epoch and found no evidence of multiplicity in images having spatial scales of 0.13"; however \citet{forveille2005}
 found clear evidence of the binary nature of the system using adaptive optics systems. The separation of the
two stars was 0.125$\pm$0.01 arcseconds during the binarity discovery epoch; \citet{gatewood2005}
 measured a parallax of 67.6$\pm$2.0 mas for a distance estimate of 14.8$\pm$0.4 pc. 
Based on this value, its M$_{J}$ is 9.76.  \citet{forveille2005} estimated an orbital
period of $\sim$5 years. The infrared colors of \citet{forveille2005} ($M_{K'}$=10.46  for the companion, and $\Delta m_{K'}=$0.26$\pm$0.05)
 lead to acceptable spectral type pairings of M7.5V+M8.5V or M8V+M9V. 
\citet{phanbao2007}
reported a single frequency detection at radio wavelengths from the binary; our team selected this object for follow-up observations at multiple wavelengths in order
to characterize the nature of the radio emission and compare with behaviors displayed by other solar neighborhood ultracool dwarfs. 
Recent high resolution spectroscopic observations with NIRSPEC on the Keck~II telescope have 
determined a combined $v\sin i$ of the system to be $\sim$50 km s$^{-1}$ (M. Rosa Zapatero Osorio, private communication).
Because of the small angular separation of the two stars, the spectrum obtained includes the contribution of both
stars.  The large $v\sin i$ measured thus could be due to the velocity separation of the two stars.
However, if we assume that the estimated $\sim$ 5 year period orbit is circular, then the small masses of the two components \citep[$\approx$0.09M$_{\odot}$;][]{burgasser2007}
implies that there would be 
a maximum orbital velocity amplitude of $\sim$20 km s$^{-1}$, and the excess velocity width could be due to
rapid rotation.  Thus it may be possible that
one or both of the two stars is rapidly rotating.
Table~\ref{tbl:obs} lists the image sensitivities of previously reported radio observations.

\subsection{2MASS J05233822-140322}
This likely brown dwarf (hereafter designated 2M0523)
was discovered by \citet{cruz2003} and given an L2.5 spectral type. 
The derived distance was 13.4$\pm$1.1 pc, implying $M_{J}$ of 12.45.  The distance is an estimate based on the
spectral type and $M_J$ calibration in the Cruz et al. paper.
Photometric variations in the I band were probed by \citet{koen2005};
the mean magnitudes in 
two runs $\sim$3 days apart agreed to within 0.01 mag, thereby providing no
indication of infrared photometric variability.
The first report on radio emission from this object was from \citet{berger2006},
with a detection of 231$\pm$13 $\mu$Jy on one out of 3 occasions, and nondetections on the other two.  
Recently, \citet{antonova2007}
described additional single frequency radio observations which failed to detect 
the object at a level which is more than a factor of 10 lower than the detected flux density.
Table~\ref{tbl:obs} lists the image sensitivities of previously reported radio observations.

\subsection{2MASS J14563831-2809463=LHS~3003}
LHS~3003 is a very low mass star with spectral type M7, at a distance
of 6.56$\pm$0.15 pc \citep{reidcruz2002}; its M$_{J}$ is 11.00.  
\citet{fuhrmeister2005} give a $v\sin i=$6$\pm$1.5 km s$^{-1}$,
noting asymmetric profiles of the H$\alpha$ and H$\beta$ emission lines,
and a change in Balmer lines over several averaged spectra (without any photometric indication 
of flaring) as evidence for numerous active regions on the surface.
\citet{fleming1993} did not detect the source in ROSAT All-Sky Survey data,
with an upper limit to the X-ray luminosity of $\log L_{X} (erg s^{-1}) \le $25.66.
\citet{schmitt1995} did detect the source with the ROSAT All-Sky Survey data
with $\log L_{X} (erg s^{-1})$ of 26.27 \citep[with apparently the same data set as ][]{fleming1993}.
\citet{mohantybasri2003} give $v \sin i =$8 km s$^{-1}$, $\log L_{\rm H\alpha}/$L$_{\rm bol}$=-4.31.
\citet{basrireiners2006} looked at radial velocities to determine if this was a 
spectroscopic binary, and found no apparent variations in radial velocity.
\citet{reinersbasri2007} detected evidence for substantial large-scale 
magnetic field strengths on the surface, finding the product of magnetic field strength and filling factor 
to be $Bf$=1.6$\pm$0.2 kG.
\citet{burgasserputman2005} reported a detection of LHS~3003 at a wavelength
of 6 cm in $\sim$ 11 hours with the Australia Telescope Compact Array (ATCA)
at a flux density of 270$\pm$40 $\mu$Jy, 
while simultaneously obtained data at 3.6 cm with the same image noise
did not reveal a detection.
Table~\ref{tbl:obs} lists the image sensitivities of previously reported radio observations.

\section{Observations and Data Reduction\label{analsec}}
Our 2006 VLA\footnote{The National Radio Astronomy Observatory is a facility of the National
Science Foundation operated under cooperative agreement by Associated Universities, Inc.}
B-configuration observations were meant to further characterize the nature of the
radio emission from LHS~3003 and 2M0523.  In order to do this, we observed in two subarrays, one 
operating at 3.6 cm (4.8 GHz) and the other cycling between 6 and 20 cm (4.8, 1.4 GHz, respectively).  
%These observations were made including 4 EVLA antennas, and possible problems
%have been reported
%in VLA-EVLA antenna pairs\footnote{See discussion
%at http://www.vla.nrao.edu/astro/guides/evlareturn/}.  We searched for these effects in our data,
%and did not find anything amiss. 
%In addition we removed EVLA antennas
%from the data, with no change in the result of no detected program sources.  
%Therefore we can rule out 
%problems associated with incorporating EVLA antennas as factors in explaining
We also extracted archival observations of LHS~3003 at 6 and 20 cm from the VLA archive and reduced those.
C-configuration
observations of LP349-25 in January 2007 were performed in two subarrays, each subarray recording one of two frequencies (3.6 or 6 cm). 
Additional A-configuration observations in August 2007 spanned roughly 20 minutes at each frequency, with data taken using the full configuration.
All data were obtained in standard continuum mode with 50 MHz bandwidth.
%This ancillary dataset was meant to probe the constancy of the radio flux from LP349-25 prior to submitting a proposal for
%higher resolution spatial imaging of the binary (see \S 3.3).
%For this dataset we received a notice that data taken in Aug. 2007 had problems with two antennas recording incorrect
%nominal sensitivities and system temperatures, so we corrected the weights of those antennas. 

After initial
data reduction and imaging, no sources were found at the location of either LHS~3003 or 2M0523 at any frequency.
We investigated and ruled out problems associated with EVLA antennas as factors in explaining the lack of detection.
%the lack of detection of 2M0523 and LHS~3003 on 30 June 2006.
Correspondence of background sources in multiple epochs also lends reliability to
the calibration and imaging steps.
%There was no detection at this earlier epoch of LHS~3003 either.
Table~\ref{tbl:obs} compares the sensitivities of observations at all epochs for the three program sources, 
including other published radio observations.  
For 2M0523 and LHS~3003, the 1$\sigma$
rms values from our observations are comparable to or less than those in the detection epochs 
at the corresponding frequency and should have detected the source if it
were emitting at the same level as during the detection epoch.
Our 3 $\sigma$ upper limits for the intensities of 2M0523 and LHS~3003 are listed
in Table~\ref{tbl:fluxes}.

LP349-25 was detected
at both frequencies at both epochs. 
%while our single epoch multi-frequency observations of LHS~3003 and 2M0523 did not result
%in any detections of our program sources.  
Maps of total intensity and circular polarization
(Stokes V parameter) were made for the
C- and A-configuration observations of LP349-25, 
first by including all visibilities and later after removing the contribution to the visibilities of background radio
sources. Single Gaussian fits to the source position near phase center were done to determine position and peak flux density. Direct Fourier
Transforms of the visibilities at the position of the source were done to examine variability (AIPS task DFTPL).
Table~\ref{tbl:fluxes} lists the average radio flux density measurements or upper limits at each epoch and wavelength for our program sources.
There was no significant level of circular polarization measured at any epoch, with  3 $\sigma$ upper limits of
78, 57 $\mu$Jy at 3.6 and 6 cm in the C-configuration observation of LP349-25, respectively, and 120, 159 $\mu$Jy at 3.6,
6 cm in A-configuration, respectively. 

\section{Data Analysis and Results}
In the following sections we describe the temporal and multi-frequency radio behavior for all three objects, in 
addition to the spatial resolution information provided by A-configuration observations of LP 349-25. 

\subsection{Temporal Behavior}
The radio emission from LP349-25 shows steady levels over long timescales.  
%The stability of the radio-emitting corona on long
%timescales 
This is revealed by the similar flux densities in 3 epochs at 3.6 cm
(Tables~\ref{tbl:sources} and ~\ref{tbl:fluxes}), identical to within 3$\sigma$ based
on differing exposure times and beamsizes.
The timespan of these 3.6 cm observations is 1.6 years.
The 6 cm flux densities measured at two epochs are also consistent to within $<$1 $\sigma$,
over a timespan of 0.6 years.
%The peak flux densities are slightly lower, but with a higher rms due to the shorter integrations; the observation-integrated
%fluxes are still consistent with the C-configuration observations and those previously reported by
%\citet{phanbao2007} at the 3$\sigma$ level at 3.6 cm. 

%The average 3.6 cm flux level of LP349-25 in the C-configuration epoch was the same, to within the error bars, as that previously determined
%by \citet{phanbao2007}.
%Furthermore, neither the 3.6 nor 6 cm light curves were consistent with any obvious variability, periodic or otherwise.
The intra-epoch behavior of LP 349-25 likewise reveals no detectable short-term variability.
We investigated this using different techniques, designed to look for periodic signals and stochastic variability,
either of which which may appear at either frequency independently.
%rules out short-term variability on LP349-25.
We performed Lomb normalized periodograms on light curves of C-configuration data made at 10 s time resolution
on both the 3.6 and 6 cm data to see if there was any evidence of periodic behavior in the radio emission, and found 
no statistically significant periodic signals.
The time span of the data analysed was 10.7 hours, with a maximum measurable period of 42 hours.
%We also examined evidence for short-term variability, computing the K-S statistic comparing the observed
%cumulative distribution of events in a light curve with binning $t_{\rm bin}$ against that expected from a 
%uniform distribution
%of events.
%For $t_{\rm bin}$ values of 10, 60, 300 and 600 seconds, at both 3.6 and 6 cm, we could not reject the null hypothesis
%that the events were uniformly distributed, and therefore no intrinsic variability was occurring.
%There was no evidence for short-term bursting
%behavior during the observations made in August 2007 during A-configuration.

\begin{figure}[htbp]
\includegraphics[scale=0.7,angle=90]{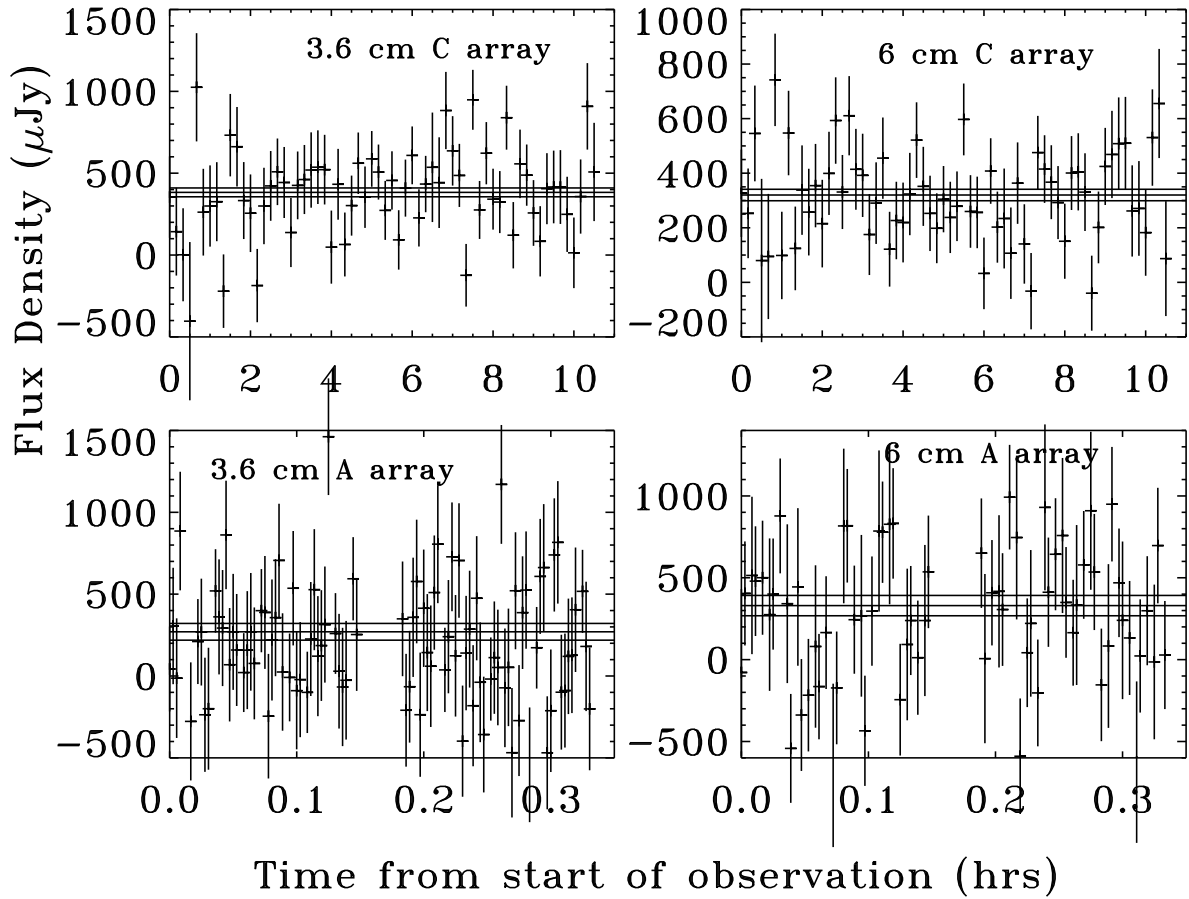}
\caption{Detail of flux density variations for LP 349-25 from two different observing epochs, at 3.6 cm (8.4 GHz) and 6 cm (8.4 GHz).
The solid line gives the average flux density and 1$\sigma$ deviations; the crosses
give individual measurements and error bars.  
For C-configuration observations the time binning is 10 minutes, and for A-configuration it is 10 seconds.
No bursting or periodic behavior is detectable.
\label{fig:lc}}
\end{figure}

We also examined evidence for short-term aperiodic variability in both the C- and A-configuration observations of LP349-25. 
We computed the chi-squared value for the fit between the
average flux density measured at 3.6 and 6 cm over the entire time interval, and the flux densities measured in finer
time bins (10, 60, 300 and 600 seonds for the C-configuration observation, and 10 seconds for A-configuration observations).
In all cases the finer flux density variations were consistent with the average flux density.
Figure~\ref{fig:lc} shows the flux density variability both between and within epochs for the C- and A-configuration observations of LP349-25.

The upper limits on radio emission from 2M0523 and LHS~3003 reflect variations of factors of 4 and 1.7, respectively,
compared to the previously detected levels of radio emission.  While the limits described in \citet{antonova2007}  
for 2M0523 are much deeper, the addition of this extra epoch at which the radio emission is undetected lowers the fraction 
of time this object produces detectable levels of radio emission, to $\sim$10\%.
The previously described detected emission from LHS~3003, in \citet{burgasserputman2005}, was accumulated over
$\sim$11 hours of integration, so the addition of this short epoch described here does not significantly affect the computed duty cycle
of $\sim$87\%.  Further observations of this interesting source are required to study the transient nature of
its radio emission.

\subsection{Multi-Frequency Behavior}
The detection epochs of LP349-25 and 2M0523 corresponded to only one wavelength (3.6 cm), and
were reported in \citet{phanbao2007} and \citet{berger2006}, respectively.  LHS~3003
was detected at 6 cm but not at simultaneously obtained 3.6 cm observations with the same sensitivity.
While further multi-frequency
observations of 2M0523 failed to detect the source at any wavelength, we have detected LP349-25 at both 
6 and 3.6 cm on two occasions.  
Thus, only LP349-25 and LHS~3003 have constraints on the multi-frequency behavior of their radio emission. 
For radio flux density $S_{\nu}$, $S_{\nu}\propto \nu^{\alpha}$, the 6--3.6 cm spectral index
computed for LP349-25 from time-averaged C-configuration observations was
 $\alpha$=0.33$\pm$0.17, and for the A-configuration observations it was
$\alpha$=$-0.47\pm$0.38.  We also computed $\alpha$ for time intervals within the C-configuration
observation corresponding to the same time bins as done in \S 4.1 (10, 60, 300, and 600 seconds)
and found no variability in $\alpha$ on these timescales.
\citet{burgasserputman2005} reported the upper limit for the radio
spectral index between 3.6 and 6 cm for LHS~3003 to be $\alpha \le$ -1.2.
%The datasets described here covered 3.6, 6, and (20 cm for 2M0523 and LHS~3003) taken in subarray modes to be sensitive
%to both single-frequency temporal behavior as well as  multi-frequency behavior.  This enables us to examine
%the bandwidth of the emitting phenomena.  
We note that for both LP349-25 and LHS~3003, the sensitivity
in each frequency band for multi-frequency observations is  very similar (numbers are listed in Table~\ref{tbl:obs}).
In addition, the multi-frequency observations were made simultaneously in the case of LHS~3003 at the ATCA (both receivers
operate at the same time),
and the 12 Jan. 2007 epoch for LP349-25.  The 15 Aug. 2007 epoch data of LP349-25 were obtained
by alternating frequencies using the full array during the same observation session, so the multi-frequency information is not
strictly simultaneous. 
Given the lack of variability on timescales $\ge$ 10 seconds described in \S4.1, however, we can consider the
multi-frequency recordings of LP 349-25 during the 15 Aug. epoch to be simultaneous.

For LP349-25, the
emission is stable at both frequencies in addition to
being stable in time.  This is revealed by the similar values returned for the radio spectral index at two epochs
together with the lack of both short-term and long-term variability.  The frequency difference between the 
3.6 and 6 cm VLA receivers is $\Delta \nu=$3.6 GHz, implying that the detected radio emission extends
across this frequency range.  
In contrast, the emission from LHS~3003 reported in \citet{burgasserputman2005} was detected at only one wavelength
at one epoch.
This implies that the emission occurs over a range in frequency $\Delta \nu \le$3.6 GHz and possibly much 
smaller. 
Unfortunately, since the multi-frequency observations of LHS~3003 we performed
on 30 June 2006 did not detect the source at any frequency, we cannot constrain the multi-frequency behavior
of LHS~3003 during this epoch. 

\subsection{Imaging of LP349-25}
\begin{figure}[htbp]
\includegraphics[scale=0.5,angle=-90]{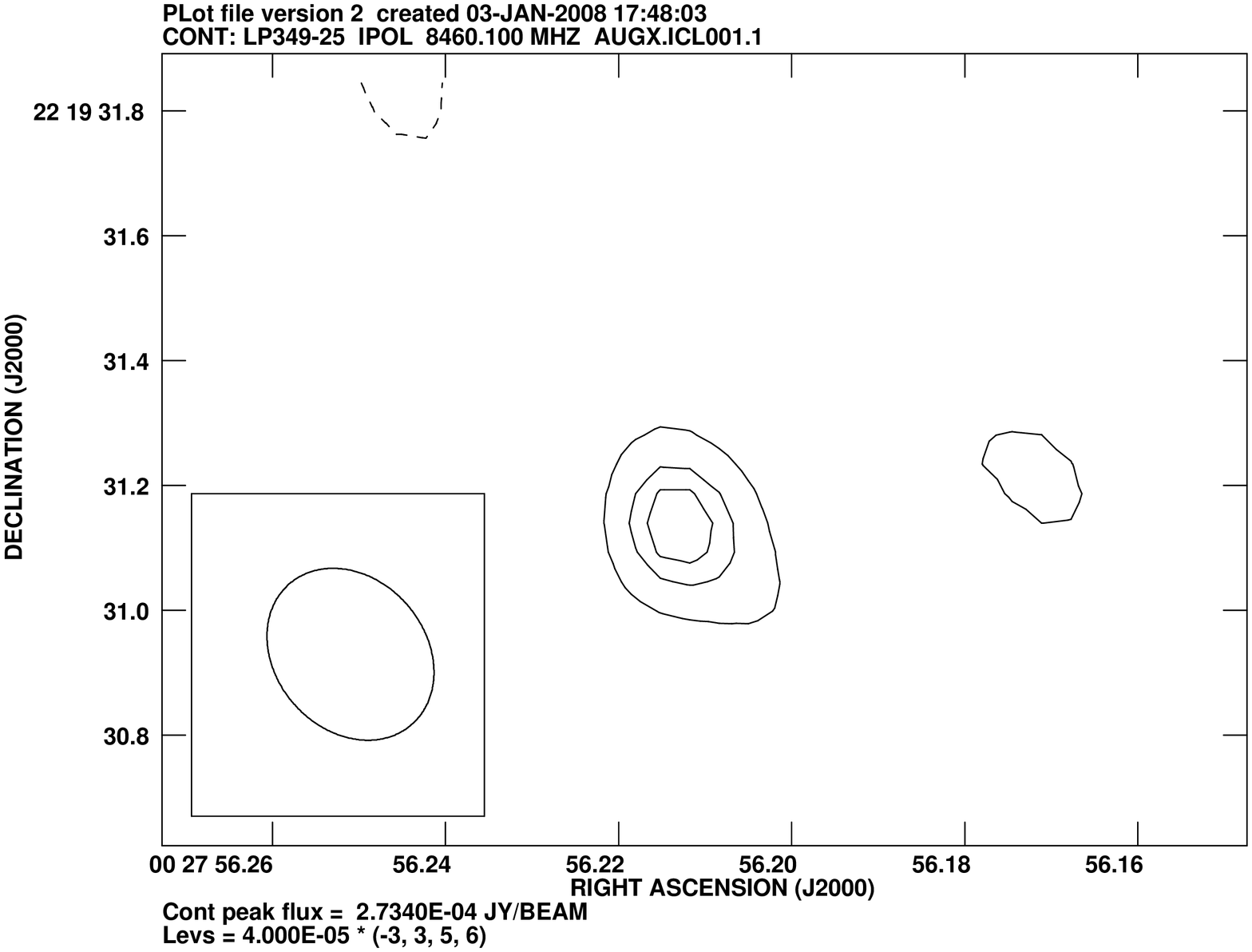}
\caption{}{A Array VLA observations of the binary LP 349-25.  The size of the radio beam is shown in the lower
left corner.  
The 1$\sigma$ rms from a blank region of the image is 40$\mu$Jy.
The peak SNR in the source is only $\sim$6.7; it is not clear whether one or both
dwarfs contributes to the radio emission. 
Details of the fit to the image contours are described in \S 4.3.
\label{fig:Aarray}
}
\end{figure}

We explored whether the contribution of both dwarfs to the radio emission from LP349-25 could be constrained
from the observation made in A-configuration.
If the two dwarfs were at the
separation noted during the discovery epoch \citep[0.125$\pm$0.01";][]{forveille2005} and had nearly equal intensity, 
we might have been able to separate the two components from our image of the A-configuration 3.6 cm data, 
made with 0.047" pixel size.
Two-dimensional elliptical Gaussian fits to the naturally weighted 3.6 cm A-configuration observations of LP349-25 return a 
major and minor axis of 0.36$\pm$0.06 and 0.23 $\pm$0.04 arcsec, with a position angle measured north from east
of 36$\pm$14 degrees.  Deconvolving the fit from the clean beam, the nominal major and minor axes of the elliptical Gaussian
are 0.21'' (minimum 0.12", maximum 0.28") 
and 0'' (minimum 0", maximum 0.09"), with a position angle of 32$^{\circ}$ (minimum 1$^{\circ}$, 
maximum 67$^{\circ}$). Minimum and maximum values are the extrema obtained by a deconvolution
of the source beam from the clean beam with 0.7 times the error of component values.
%{\it The minimum and maximum values are obtained by deconvolving the source beam parameters
%with all combinations of 0.7* error and listing the extreme values.}
Figure~\ref{fig:Aarray} shows a close-up of the radio contours.  
We explored using uniform or ``super-uniform'' weighting to 
enhance the longest baselines compared to the shortest ones, however these weighting schemes
also increase the noise and these images produced peaks at lower significance value, with evidence for
only one source at the location of LP349-25; there was no evidence for a second source out to the maximum distance sampled
in the A-configuration image, 24" from LP349-25.
This is perhaps not surprising:
The discovery epoch on July 3, 2004 of the binarity of LP349-25 \citep{forveille2005} revealed the separation between
the two stars to be 0.125$\pm$0.01" at a position angle of 12.7$\pm$2.0$^{\circ}$
east from north, while an earlier epoch on Sept. 18-19, 2001 \citep{close2002} showed the system to be
unresolved.  This earlier epoch occurred
almost 6 years prior to our observations, and with the estimate by \citet{forveille2005}
that the orbital period is about 5 years an explanation for why the two stars were not resolved is apparent.
The A-configuration observations thus do not constrain whether one or both dwarfs in the LP349-25 system contribute
to the total observed radio emission; further observations at higher spatial resolution are needed to resolve the two dwarfs.

Spatially resolving the two stars in the binary system is important for two reasons.  The first is that 
this will allow us to determine the relative amount of radio emission arising from each dwarf, and hence
explore radio emission mechanisms further.  As noted by \citet{phanbao2007}, the radio luminosity from the system
is slightly higher than the steady levels seen from other radio-emitting ultracool dwarfs.
The second lies in the unique nature of LP349-25 as a nearby tight binary with an orbital period
amenable to astrometric monitoring; direct mass measurements are crucial to placing
constraints on the formation processes of very low mass ($<$ 0.1 M$_{\odot}$) stellar objects
\citep{burgasser2007}.
Mass determinations are also important for comparing with theoretically derived masses from evolutionary models, and
establishing mass-luminosity relations in this mass range \citep{bouy2004}.  
Due to the low flux density, sensitive high spatial resolution observations such as those attainable
with the High Sensitivity Array are needed; we have an approved program to begin this monitoring, which should 
be able to attain an astrometric precision of $\approx$ 30 $\mu$as using phase referenced
observations \citep{pradel2006}.
LP 349-25 thus provides the keys to understanding the
formation of ultracool dwarfs as well as the enigmatic nature of their radio emission.

\section{Discussion}
Of our observations of three ultracool dwarfs, the detections of LP 349-25 offer the most 
stringent constraints on the types of radio emission produced in ultracool dwarfs.
The relevant and interesting facts about the radio behavior seen on LP 349-25 are
that the emission is constant on both short- and long timescales, and that the
observed multi-frequency radio emission has a flat spectral index which also does not
appear to vary on short and long timescales.  
This is unusual behavior in light of the fact that the few ultracool dwarfs which have been
detected at radio wavelengths have all exhibited dramatic variability, whether periodic
or bursting behavior.
Here we explore
what constraints these details provide within the context of behavior seen on 
active stars and ultracool dwarfs.

\subsection{Constraints from Absence of Short-Term Variability}
The lack of short-timescale variability on LP349-25 on timescales of 10 s or larger
is constrained by the observations conducted
on 12 Jan. 2007, for 10.7 hours at both 3.6 and 6 cm.
This 
observation duration should have been long enough to discern periodic variability:
for $v\sin i \gtrsim$10 km s$^{-1}$, the rotation period for a 0.12 R$_{\odot}$ star of mass $\sim$0.09 M$_{\odot}$ \citep{pont2005} is $<$ 10 hours (in which case
we should have seen one complete rotation period), and
for v$\sin i \gtrsim$20 km s$^{-1}$, we should have seen two or more rotation periods.
Rotational modulation of the radio emission from ultracool dwarfs is a commonly observed behavior
\citep{hallinan2006,hallinan2007,hallinan2008}, and so its lack in the case of LP349-25 is significant.  
We note that the greatest constraints on lack of short-term variability come from our C-configuration
observation spanning 10.7 hours, but the observation of LP349-25 described by \citet{phanbao2007}
lasted 1.7 hours.
Expressed as a fraction of the unknown rotation period this span $\Delta t$ is 1.7 hours/P;
taking the upper limit on P to be 10 hours, this observation spanned a fraction
of the rotation period $\Delta P/P>0.17$.  For
short periods (2--3 hours), this is a significant fraction of the rotation period, and
the lack of variability seen during this observation provides an additional constraint
on the lack of rotationally modulated for the case of short periods.

The absence of variability in LP 349-25 on timescales comparable to the 
%There are several possibilities which could explain the lack of variability on timescales comparable
rotation period of one or both stars imposes spatial constraints.
The viewing orientation could explain the lack of variability:
if the inclination of the binary system is situated close to pole-on 
then no modulation of the radio emission is to be expected if either or both dwarfs produces radio
emission.
However, the preliminary measurement of a substantial $v\sin i$ discussed in \S2.1 suggests that one or both of the dwarfs
may be rapidly rotating, and for a pole-on configuration no rotational broadening of emission lines
is expected.
%The emission could be anisotropic arising from an isolated structure, 
%in which case a favorable viewing geometry could explain
%the lack of short-term variability; if our line of sight
%always intersects the opening angle of the emission, no variability over the course of a rotation period
%would be seen.  
However, if the system were not observed at a nearly pole-on configuration, then the pattern of temporal emission would 
resemble the pulsar behavior seen in other ultracool dwarfs \citep{hallinan2006,hallinan2007,hallinan2008}.
There could be many emitting structures 
whose properties
are uniform and so cause a lack of observed rotational modulation.  
This would require a high degree of 
homogeneity of the magnetic structures over the surface of one or both
stars.  A peculiar arrangement could
be obtained between both ultracool dwarfs
where both are radio-emitting but do so out of phase with each other, so that the total observed
radio emission appears to be constant.  This seems unlikely, as it would require rather special conditions
to occur; namely, the two dwarfs are tidally locked or information can be communicated on timescales
$<$10 seconds. The separation of the two dwarfs at the discovery epoch was $\approx$1.8 AU, and this distance
is too large for tidal locking to occur: the synchronization timescale given by \citet{zahn1977} is $\sim$34 GYr. 
Finally, a circumbinary radio-emitting structure could be present, but given the separation of the two dwarfs
($\gtrsim 3000$ R$_{\star}$ at the discovery epoch)
this is also unlikely.

Since variability is a key factor controlling the observed radio emission from other
ultracool dwarfs, we explore some physical constraints imposed by the variability
timescales here.  Timescales as short as a few minutes have been observed during impulsive
bursting behavior \citep{burgasserputman2005,berger2008}.
We explore the time evolution of the radio emission on short timescales
under the constraints that no emission variability is observed.
If the emission is due to singular injection events akin to the isolated bursts seen in other
systems, then assuming that the emission region is localized,
%Assuming that the emission region is localized,
energy loss can occur through collisional losses in a high
density environment, or radiation losses in a high magnetic field region.
The timescale for this energy loss then can be related to the limits on observed
variability, to provide a constraint on the plasma environment.
%Either limit argues for an extreme plasma environment. 
The collisional deflection time for a
particle of energy $E_{\rm kev}$ is \\
\begin{equation}
t_{c} = 9.5\times 10^{7} \frac{E_{kev}}{n_{e}} \frac{20}{\ln (4.2\times10^{5} T/\sqrt{n_{e}})} \;\;\; s
\end{equation} 
\citep[eq. 2.6.20 of][]{benz2002}, with $E_{\rm kev}$ the particle energy in keV, electron density $n_{e}$ in
cm$^{-3}$ and temperature $T$ in K; for T$>$10$^{6}$K  and a 10 keV particle
to have t$_{c} <$10 s requires $n_{e}\gtrsim$ 
3$\times10^{8}$
cm$^{-3}$, while t$_{c} >$10.7 hrs requires n$_{e}\lesssim$ 4$\times$10$^{5}$ cm$^{-3}$.
The constraints do not depend tightly on the choice of T.
The timescale for radiation loss is \\
\begin{equation}
t_{r} = \frac{6.7\times10^{8}}{B^{2}\gamma} \;\;\; s
\end{equation}
\citep{petrosian1985}
where $B$ is the magnetic field strength in Gauss, $\gamma$ the Lorentz factor;
for a 10 keV electron, t$_{r}$ is greater than 10.7 hours for $B \lesssim$130 G or less than 10 s for $B\gtrsim 10^{4}$G. 
%Collisional losses will dominate in a high density, low magnetic field strength region, primarily at
%low energies, and radiation losses will dominate in a low density, high magnetic field strength region,
%primarily at high energies. 
Although we cannot rule out that variability is occurring 
on timescales less than 10 seconds, the extreme plasma conditions
($n_{e}\gtrsim$ 3$\times10^{8}$ cm$^{-3}$ or $B\gtrsim 10^{4}$G) make the scenario of rapid variability
on these timescales implausible.
Therefore we consider that the observations constrain the plasma conditions to be either $B\lesssim$ 130 G
or $n_{e} \lesssim$ 4$\times$10$^{5}$ cm$^{-3}$.

\subsection{Constraints from Stability over Long Timescales}
The lack of long-timescale variability on LP349-25 is constrained by the three epochs of flux density measurements
at 3.6 cm, and two epochs of measurements at 6 cm.  
The average flux densities at 3.6 cm are 5\% and 30\% different comparing the Dec. 2005/Jan. 2007 and
Aug. 2007/Jan. 2007 flux density measurements, respectively,
while at 6 cm the average flux densities are 6\% different.
%The fact that the flux density measurements 
%are statistically identical 
Even in active stars with 
multiple measurements of ``quiescence'' outside of any obvious variability, different radio flux densities
and spectral indices
are obtained \citep[see Figure~3 and discussion in][]{gudel2002}, revealing differing radio-emitting magnetic configurations.
Table~2 of \citet{ostenetal2004} shows a factor of fifteen variation in the average 3.6 cm flux density
of the active binary HR~1099, and \citet{caillault1988} noted that 6 cm detections and upper
limits for different epochs of radio observations of BY~Draconis variables varied by as much as
an order of magnitude.
The rotational modulation noted by \citet{lim1992}
of AB Dor's radio emission is only sporadic,
indicating a lack of long-term stability or repeatability of the flux density.
The well-studied radio-emitting ultracool dwarf TVLM513-46546 also displays different radio emission properties
over long timescales, ranging from variability of quiescent emission \citep{hallinan2006}
to periodic bursting radio emission \citep{hallinan2007} to steady quiescent emission \citep{berger2008}.
The changes in the radio emission properties in the case of both active stars and ultracool dwarfs
can be linked to the finite lifetime of the starspots or magnetic structures which are presumed to give rise 
to the radio emission.
Starspots on active stars have lifetimes  which
%\citep{spotlifetimes} 
depend on their latitude: long-lived spots lying at high latitudes can survive for years \citep{strassmeier2002}
but starspots at lower latitudes have lifetimes of less than about a  month \citep{hussain2002}.
\citet{spotlifetimes} also showed that starspot lifetime on active stars was longer for larger spots.
These trends are due to the competing effects on magnetic flux transport of differential rotation, convection,
and meridional flows.
The changes in the optical properties of the ultracool dwarf TVLM513-46456 are characterized in one epoch by
rotational modulation of isolated starspots \citep{lane2007} and in another by rotation of persistent dust clouds
\citep{littlefair2008}, confirming the finite lifetime of magnetic structures in ultracool dwarfs as well.
So the stability of the radio-emitting source in LP 349-25 over similar timescales as those during
which the radio emission from TVLM513-46546 was observed to change dramatically is unique. 
This remarkable stability
is consistent with similar conditions being
%present in the emitting structures during
the three (two) epochs when 3.6 cm (6 cm) observations were obtained, over timescales of more than a year.  
This indicates that the magnetic structure giving rise to the radio emission for LP 349-25
lies predominantly at high latitudes,
if the same magnetic structure is giving rise to the radio emission in all three epochs.
It also suggests that the lifetimes of the magnetic structures are long-lived, or alternatively
that there is little
variation in the properties of the magnetic structures which form.

\subsection{Constraints from Spectral Index}
The spectral index measurements of LP 349-25 during the C- and A-configuration epochs
described in Section 4.2 are the same to within 2 $\sigma$, and
consistent to 2 $\sigma$ with a value of 0. 
For these purposes we consider this ``flat'' even though with the same statistical significance 
the two spectral 
index measurements are compatible with the range $-0.01<\alpha<0.29$.
Section 4.2 also described the lack of evidence for changes in the spectral index
within the C-configuration observations on timescales greater than 10 seconds.
A primary constraint on the radio emission mechanism from the spectral index
is that it does not change significantly.  At the most basic level, this indicates that there were no
observed changes in optical depth on timescales either within the C-configuration
observations ($>$ 10 seconds), or on longer timescales between the two epochs in which
both 3.6 and 6 cm observations were obtained (0.6 years). 

Optically thin thermal bremsstrahlung naturally produces a flat spectral index \citep{dulk1985}.
In this case, the optical depth is expressed as $\kappa_{\nu} \sim$0.2 $n_{e}^{2} L/\nu^{2}/T^{3/2}$,
where $T$ is the temperature of the thermal plasma, $L$ is the characteristic length scale 
along the line of sight, and $n_{e}$ and $\nu$ are the electron density and observing frequency,
respectively. 
The brightness temperature T$_{b}$=T$\tau$ for $\tau \ll$1, and the 
the flux density S$_\nu$ is \\
\begin{equation}
S_\nu= \frac{k_{B}\nu^{2}}{c^{2}} T_{b} \Delta \Omega\;\;\; ,
\end{equation}
where k$_{B}$ is Boltzmann's constant, $c$ the speed of light, and $\Delta \Omega$ the differential
solid angle. 
Writing $\Delta \Omega$ as $A/d^2$ with $A$ the source area and $d$ the distance to the target,
we can express the flux density as a function of the observed temperature and volume emission measure
($VEM=n_{e}^{2}\times A \times L$), \\
\begin{equation}
S_\nu = 1.4\times10^{-54} \frac{VEM}{\sqrt{T}} \; \;\; Jy
\end{equation}
where the volume emission measure has units of cm$^{-3}$ and temperature is in units of K.
Thermal bremsstrahlung emission from a chromosphere or corona at temperatures from 10$^{4}$ -- 10$^{6}$K
imply volume emission measures around 10$^{53}$ cm$^{-3}$ for the observed levels of
radio flux density.

As described in \citet{ostenetal2006}, the distribution of spectral indices for a sample of
active K and M dwarfs taken from \citet{gb1996} has an average of $-0.4$ and FWHM of 1.5.  
The behavior of LP349-25 is consistent with this
spread.  
The interpretation for active stars is that the emission arises from an inhomogeneous 
source emitting optically thin gyrosynchrotron emission.
In the optically thin limit, for a value
$\delta$ parameterizing the number density distribution of accelerated electron energies ($N(E) \propto E^{-\delta}$), the flux density depends on frequency as 
S$_{\nu} \propto \nu^{1.22-0.9\delta}$, so
a flat spectrum could be obtained if the spectrum of nonthermal electrons is 
particularly hard, $\delta \sim 1.3$.

Another possibility to explain a flat spectral index is a hybrid of the above two models, 
optically thick gyrosynchrotron emission 
with opacity provided by collisions.  This model obtains a flat spectral index 
for a value $\delta$ of 3.6.
For this model, a large electron density is
required for the collisional opacity to exceed the gyrosynchrotron opacity.  
For the conditions described in \S 5.1, namely T$\sim$10$^{6}$K and B$\lesssim$130 G,
$n_{e}$  must be $\gtrsim$ 10$^{14}$ cm$^{-3}$ for values of the total nonthermal electron
density $>$ 1.
This is at odds with the restrictions in \S 5.1 necessitating a low electron
density  ($n_{e} \le 4\times$10$^{5}$ cm$^{-3}$)
for the plasma cooling time to be long enough that no variability would have
been seen.
%Using equations in \citet{dulk1985} we can evaluate the bremsstrahlung and gyrosynchrotron opacities
%given the constraints on electron density and magnetic field
%strength in the source derived in \S 5.1 on the plasma environment given the lack of 
%short-timescale aperiodic variability, and find that for a range of reasonable values of 
%the plasma temperature and number density of accelerated electrons, the 
%pulsar behavior.  A nearly pole-on configuration of a stable beamed
%source could explain the steady nature of the 
%radio emission on short timescales, but requires no rotational broadening.

\subsection{Putting It All Together: Implications for the Radio Emission in LP349-25}
The discussion in the above three sections describes temporal and spectral index constraints
on the radio emission from LP 349-25.  Here we examine these constraints, together with the
upper limits on circular polarization and possible values of brightness temperature, for what they
can tell us about the radio emission from this ultracool dwarf binary.

The simplest explanation which fits the lack of variability and flat spectral index
observed on LP 349-25 is that the emission arises from optically thin thermal bremsstrahlung
radiation distributed uniformly across one or both of the stellar disks.  
However, this model runs into several problems.  
If we assume that this material is distributed homogeneously across the disk of one star, 
with length
scale equal to R$_{\star}$ or smaller, then electron densities in excess of 2.2$\times$10$^{11}$ cm$^{-3}$ are
required to produce the volume emission measure determined in \S 5.3.
The plasma frequency corresponding to such a density n$_{e}$ would be 
$\nu_{p} \gtrsim$ 4.2 (n$_{e}$/2.2$\times$10$^{11}$)
 GHz, with n$_{e}$ in cm$^{-3}$, and we would expect a cutoff at or below the 6 cm observing frequency.
Optically thin radio bremsstrahlung emission would also show up at longer wavelengths as X-ray emission,
and given the distance of LP349-25, implies an X-ray flux from 0.2--2.4 keV of $\sim$4$\times$10$^{-11}$ erg cm$^{-2}$, or L$_{X}$ of 10$^{30}$ erg s$^{-1}$.  
This would put LP349-25 in the top 101 brightest X-ray stars within 50 pc 
\citep{xraystars} and would make the lack of a detection in the ROSAT All-Sky Survey,
which had sensitivities well below this flux value \citep[down to 10$^{-13}$ erg cm$^{-2}$ s$^{-1}$;][]{rass}
 problematic.  It would equally well
represent a discontinuity with the X-ray behavior of other ultracool dwarfs, which have $L_{X}$
typically around 10$^{26}$ erg s$^{-1}$ \citep{stelzer2006} when they are detected.
%If the emission does indeed arise from a homogeneous optically thin thermal source, then
%in the presence of a magnetic field, polarization in the sense of the extraordinary mode is expected 
%\citep{dulk1985}, although the limits from our observations.

The emission from radio-active stars and ultracool dwarfs by extension is usually
associated with the presence of enhanced magnetic fields, which naturally leads to
gyrosynchrotron or synchrotron emission for incoherent radiation, or 
cyclotron maser emission for coherent radiation.  These mechanisms require accelerated particles,
and so the constraints on the plasma environment deduced in \S 4.1 from the lack of aperiodic
short-term variability are applicable here.
The general gyrosynchrotron model as applied to active stars
would require a spatially inhomogeneous source to arrive 
at the flat spectral index.  However, the lack of short-term 
variability described in \S4.1 require that the same emitting
region be visible over the one or more rotation periods seen in the C-configuration data.
Based on expected lifetimes of starspots seen in active stars and the evidence for
epoch-to-epoch changes of radio emission in another well-studied ultracool dwarf,
a long-lived polar spot and the binary having a high orbital/rotationl inclination are plausible. 
This would naturally explain both the lack of short-term and long-term variability,
as the high inclination would lead to no rotational modulation over the short term,
and the similar values of flux density and spectral index are consistent with the same
emitting structure producing the observed radio emission.  A highly homogeneous source
with a hard distribution of electrons would also produce a flat spectral index, as discussed
in \S 5.3.
Either of these explanations requires that there is little
long-term evolution of magnetic structures responsible for producing the radio emission,
which is unique amongst active stars.  It is also at odds with the preliminary
evidence that one or more of the stars produces a large $v\sin i$ signature discussed in \S2.1.
Gyrosynchtron emission usually arises at frequencies corresponding to harmonic numbers
$s$ between 10 and 100 of the electron gyrofrequency, given by 
$\nu=s \nu_{B}$, with $\nu_{B}=2.8\times$10$^{6}B$, $B$  being the magnetic field in the
radio-emitting source in Gauss and $\nu_{B}$ in Hz. 
The ranges of magnetic field strengths derived from the two observing frequencies is
17$<B<$170 G for 6 cm emission, and 30$<B<$300 G for 3.6 cm, which is compatible with
the magnetic field strength derived in \S 4.1 to explain the lack of short-term variability.
A hybrid collisional/gyrosynchrotron model, although producing a flat spectral index,
is difficult to reconcile based on the high electron density required for collisional
opacity to exceed gyrosynchrotron opacity.
The inferred brightness temperatures, as discussed in \citep{phanbao2007}, are not constraining, due to 
uncertainties in the source size.
Research on solar bursts indicates that gyrosynchrotron radiation can be directive \citep{leegary2000}, 
although
the requirements for lack of periodic variability in this case
require that the source be located at high latitudes, with a high inclination, and
either isotropic or beamed emission can be accomodated.

A coherent mechanism like electron-cyclotron maser could also be operating to produce the radio emission
from LP 349-25.  While the intense bursts observed from several ultracool dwarfs have been
interpreted using this framework, this is also problematic within the context of
the observations and constraints described here.  
The 
observed emission shows no net polarization, which
could be explained if the emission is intrinsically polarized but becomes depolarized due to 
propagation and scattering effects. 
\citet{hallinan2006} noted that the average emission from TVLM513-46546 was flat, but these measurements
were not obtained simultaneously and includes periodic emission, and so it is not clear
if the flat spectral index measured from TVLM513-46546 has any bearing on the observed flat
spectral index for LP349-25 in the absence of any variability.
Cyclotron maser emission at the fundamental or second harmonic is emitted at frequencies
$\nu=s \nu_{B}$ where $s$ is 1 or 2 respectively.
A maser operating at the observing frequencies of 4.8 (8.4) GHz requires a magnetic field
in the emitting region of 1.7 (3) kG for fundamental emission at 6 (3.6) cm, and 
0.86 (1.5) kG for second harmonic emission at the same respective frequencies.
Given the constraints on magnetic field strength from lack of variability
on short and long timescales from \S5.1, namely that B$<$ 130 G
or B$>$10$^{4}$G, in neither case would a maser would be observable at 
3.6 or 6 cm.
Plasma conditions must also be favorable for a cyclotron maser to operate, the
ratio of plasma frequency to cyclotron frequency, $\nu_{p}$/$\nu_{B}$, being less than one.
For fundamental or harmonic emission, with the constraints on $n_{E}$ and $B$
from \S 4.1, this ratio is $\sim$0.01, and so is met.
There is no general context to explain a particular value of spectral index under
electron cyclotron maser emission.
Broad-band coherent emission as suggested by \citet{hallinan2006} requires that the radiation
be emitted at multiple points in the stellar atmosphere, 
or arise from relativistic electrons $\Delta \nu/\nu \sim (\gamma-1)/\gamma$.
We can dismiss the latter case because
this implies $\gamma \approx$2, or electrons with kinetic energy $>$511 keV,
which would be much more energetic than have ever been detected from active stars; even
discussions of the electron cyclotron maser on the Sun consider particle energies of a few
tens of keV \citep{tangwu2009}.
In addition, in order for electrons of this energy to retain their energy for timescales longer than 10.7 hours
requires plasma environments of $n_{e}<$10$^{6}$ cm$^{-3}$ and B$<$90 G to survive against collisional losses and
radiative losses, respectively.
This is 
incompatible with the requirement to have the magnetic field in the emitting region 
be between 0.86--3 kG based on identification of the observing frequency with 
the fundamental or harmonic of the electron gyrofrequency.
It is always possible that the maser is continuously resupplied with relativistic 
particles but this requires that particle acceleration be steady over the long timescales
constrained here (1.6 year), and suggests an even lower magnetic field strength in the source if
the particles are to survive against radiative losses.
This may necessitate a resupply of relativistic electrons
over these long timescales with no change in intensity or spectral index.
Astrophysical sources of particle acceleration are by and large associated with 
variability, either radio or X-ray synchrotron variability, and indicate
stochastic acceleration.  
Under these conditions, we find the cyclotrom maser interpretation of this steady radio
emission from LP 349-25 to be problematic, and in light of this we conclude that an incoherent
mechanism is most compatible with the observational constraints.

%Beaming has been presented within both contexts to explain periodicities observed in stellar 
%radio emission \citep{lim1994,hallinan2006}.

In summary, the variability constraints argue for a high latitude source with a high rotational/orbital
inclination 
in order not to observe
any rotationally-induced intensity changes.  
This same source should survive for at least 0.6 yr to explain the similar values of $\alpha$ in two 
epochs, and requires that there is little evolution of the magnetic configuration that gives
rise to the radio emission.  
Gyrosynchrotron emission from a spatially inhomogeneous or homogeneous source can 
explain the flat spectral index.  The plasma environment must consist of low densities 
(n$_{e} \lesssim$4$\times$10$^{5}$ cm$^{-3}$) and moderate
magnetic field strengths ($B<130$ G) in order that the high energy electrons not lose energy
by collisions or radiation on timescales comparable to those imposed by our C-configuration
observations.  In contrast, problems are encountered with a cyclotron maser
mechanism explanation in key aspects of the observations.  
Within the context provided by the lifetimes of starspots,
and with the expectation that the radio emission arises from a region of enhanced magnetic fields,
the constraints imply that a stable polar spot emitting gyrosynchrotron emission
may be the origin of the steady radio emission from LP 349-25. 
Further monitoring of LP 349-25 to determine whether it is variable
may help answer other
questions about radio emission from ultracool dwarfs, namely the nature of
the unpolarized ``steady'' radio emission observed outside of highly polarized bursts
on other ultracool dwarfs.

Note that the overall detection rates of radio emission for ultracool dwarfs, $\sim$10\%, are not that different from the 
optical flare duty cycles of 5--7\% \citep{reid1999,martin2001} and the implied radio duty cycles of $\sim$10\%
for the transiently emitting object 2M0523.
There may be a significant bias in the the number of objects which produce detectable levels of radio emission
if variability affects the detection rates; potentially more ultracool dwarfs may be capable of producing
detectable but transient levels of radio emission than are currently assessed by single epoch surveys. 
This highlights the need for multiple epoch observations of radio-detected ultracool objects, and probably 
further monitoring of objects previously undetected in earlier epochs.
The recent detection of several unidentified submillijansky radio transients \citep{bower2007}
suggests that there may be a large reservoir of low mass or ultracool stellar objects at distances
of $\approx$ 1 kpc producing transient radio emission which can contribute to the rate of radio transients.
Further characterizing the behavior of nearby low mass and ultracool dwarfs is therefore necessary to
be disentangle a stellar component from this unknown source population.

\section{Conclusions}
Our investigation of the radio emission from three ultracool dwarfs yields useful constraints
from only one.  The temporal and multi-frequency radio observations of LP 349-25 suggest that
the radio emitting structures are located at a high latitude on one of the stars, and that the 
rotational and orbital axes have a high inclination with respect to our line of sight, 
in order to explain the lack of 
variation associated with orbital or rotational modulation.
This is potentially at odds with the preliminary determination of the $v\sin i$ of the system.
The absence of transient or aperiodic variability over short timescales also constrains
the plasma conditions which must be present, in order that high energy particles not lose
their energy through collisions or radiative losses.
The similar flux density values and spectral indices returned from epochs 0.6 years apart
are consistent with a long-lived magnetic structure giving rise to the emission.
We explore the implications for the radio-emitting mechanism, and rule out a thermal
source.  Gyrosynchrotron emission from an inhomogeneous or homogeneous source
can explain the 
flat spectral index and is compatible with the limits imposed on the plasma environment
from short-timescale constraints.
An electron-cyclotron maser is not compatible with the plasma constraints.
In terms of its observational properties and the implications provided thereby
for the radio emission mechanism, LP 349-25 is unique amongst ultracool dwarfs in exhibiting
an incoherent mechanism, and suggests that the mixed behavior of active stars
at radio wavelengths, exhibiting one or both of gyrosynchrotron and a coherent emission
mechanism, can survive to the substellar boundary.
Temporal constraints on long timescales imply that 
there is little
long-term evolution of magnetic structures, an important result for the magnetic fields
only recently established to be present in ultracool dwarfs.
Further monitoring will confirm whether LP349-25 is truly unique in
its steady multi-frequency emissions,
and interferometric observations at higher angular resolution are needed to determine whether
one or both dwarfs contributes to the total radio emission.
Both steady and transient radio emission have been presented in this paper, and our conclusions
support continued monitoring of all radio-detected ultracool dwarfs in order to elucidate
further the nature of radio emission near the substellar boundary.

\acknowledgements
This paper represents the results of VLA programs AO205, AO214, and AO223. 
Support for this work was provided by NASA through Hubble Fellowship grant \# HF-01189.01 awarded
by the Space Telescope Science Institute, which is operated by the Association of Universities for
Research in Astronomy, Inc. for NASA, under contract NAS5-26555.

%\bibliographystyle{natbib}
%\bibliography{/Users/rosten/papers/lp349-25/lp349}

\clearpage

\begin{deluxetable}{lcrcccclc}
\rotate
\tablewidth{0pt}
\tablenum{1}
\setlength{\tabcolsep}{0.03in}
\tablecolumns{9}
\tablecaption{Properties of Program Sources \label{tbl:sources}}
\tablehead{\colhead{Source} & \colhead{Spectral Type} & \colhead{Dist.} &\colhead{M$_{J}$} &\colhead{F$_{R}$} 
& \colhead{$\lambda_{\rm obs}$} &\colhead{Obsn. Date}
&\colhead{Ref. for F$_{R}$\tablenotemark{a}} & \colhead{L$_{R}$}\\
\colhead{} & \colhead{} & \colhead{(pc)} &\colhead{} & \colhead{($\mu$Jy)} & \colhead{(cm)} & \colhead{}
&\colhead{} &\colhead{(erg s$^{-1}$ Hz$^{-1}$)}
}
\startdata
LHS~3003 & M7 & 6.2& 11.0 & 270$\pm$40& 6 & 2 May 2002 & BP05 & 1.2$\times$10$^{13}$\\
LP349-25 & M7.5V+M8.5V or M8V+M9V & 14.8 &9.76\tablenotemark{b}  &365$\pm$16& 3.6 & 20 Dec. 2005 &PB07 & 9.6$\times$10$^{13}$\\
2M0523 & L2.5 & 13.4  & 12.45& 231$\pm$13& 3.6& 18 June 2004 &B06  & 5.0$\times$10$^{13}$\\ 
\enddata
\tablenotetext{a}{PB07=\citet{phanbao2007}, B06=\citet{berger2006},BP05=\citet{burgasserputman2005}}
\tablenotetext{b}{combined value}
\end{deluxetable}

% table of previous/current observations 
\begin{deluxetable}{lclll}
\tablewidth{0pt}
\tablenum{2}
\tablecolumns{5}
\tablecaption{Sensitivity of Past and Current Radio Observations of Program Sources \label{tbl:obs} }
\tablehead{\colhead{Date}  &
\colhead{$\lambda_{\rm obs}$} & \colhead{1$\sigma_{\rm rms}$\tablenotemark{a}} & \colhead{t$_{\rm exp}$} & \colhead{Ref.\tablenotemark{b}}  \\
\colhead{} &   \colhead{(cm)} &\colhead{($\mu$Jy)}  & \colhead{(hr)} &\colhead{} } 
\startdata
\multicolumn{5}{c}{--- LHS~3003 ---} \\
4 May 1994 &     6 & 70 & 0.22 & current paper\\
4 May 1994 &   20 & 590  &0.27 & current paper\\
2 May 2002 &   3.6  & 40 & 11  &BP05\\
2 May 2002 &   6  & 40 & 11  &BP05\\
30 June 2006 &  3.6 &  30& 2.6 &current paper\\
30 June 2006 & 6 &   52 & 1.4 &current paper\\
30 June 2006 & 20 &  530 &1.4 &current paper\\
\multicolumn{5}{c}{ --- LP 349-25 ---} \\
20 Dec. 2005 & 3.6 & 16 & 1.7 &PB07\\
12 Jan. 2007 & 3.6 & 27 & 11.2& current paper\\
12 Jan. 2007 &  6 & 21 & 10.7 &current paper\\
15 Aug. 2007 & 3.6 & 40 & 0.3& current paper\\
15 Aug. 2007 & 6 & 54 & 0.3 &current paper\\
\multicolumn{5}{c}{--- 2M0523 ---} \\
3 May 2004 &    3.6 &  16 &1.6  &B06\\
17 May 2004 & 3.6  &  20 &1.7 &B06\\
18 June 2004 &  3.6 &  13 &2.1 &B06\\
30 June 2006 & 3.6 &  19 &6.2 &current paper\\
30 June 2006 & 6 &  36 &3.0 &current paper\\
30 June 2006 & 20 &  118& 3.4 &current paper\\
23 Sept. 2006 &  3.6 & 15 & 9  &A07\\
\enddata
\tablenotetext{a}{$\sigma_{\rm rms}$ was determined from blank regions of naturally-weighted images.}
\tablenotetext{b}{PB07=\citet{phanbao2007}, B06=\citet{berger2006},BP05=\citet{burgasserputman2005}, A07=\citet{antonova2007}}
\end{deluxetable}

\begin{deluxetable}{lcc}
\tablewidth{0pt}
\tablenum{3}
\setlength{\tabcolsep}{0.03in}
\tablecolumns{3}
\tablecaption{Measured Flux Densities, Upper Limits for Reported Epochs \label{tbl:fluxes}}
\tablehead{\colhead{Source} & \colhead{$\lambda$} & 
\colhead{Flux Density or 3 $\sigma$ Upper Limit} \\
\colhead{} & \colhead{(cm)} &  \colhead{($\mu$Jy)}
}
\startdata
 LHS~3003\tablenotemark{a} & 3.6  &$<$ 90 \\
 "     \tablenotemark{a} & 6  &$<$ 156 \\
 "     \tablenotemark{a} & 20  &$<$ 1590 \\
 LP349-25\tablenotemark{b}& 3.6  &383$\pm$27 \\
 " 	\tablenotemark{b}    & 6    &320$\pm$21 \\
 "\tablenotemark{c}& 3.6  &262$\pm$40 \\
 "\tablenotemark{c} 	    & 6    &338$\pm$54 \\
 2M0523\tablenotemark{a} & 3.6  &$<$57 \\
 "     \tablenotemark{a} & 6 &$<$108 \\
 "    \tablenotemark{a}  & 20 &$<$ 354 \\
\enddata
\tablenotetext{a}{Epoch of 30 June 2006; see Table~\ref{tbl:obs}}
\tablenotetext{b}{Epoch of 12 Jan. 2007; see Table~\ref{tbl:obs}}
\tablenotetext{c}{Epoch of 15 Aug. 2007; see Table~\ref{tbl:obs}}
\end{deluxetable}

\clearpage

\end{document}